\documentclass[runningheads]{llncs}

\usepackage[T1]{fontenc}
\usepackage{graphicx}
\usepackage{wrapfig2}
\usepackage{amsmath,amssymb}
\usepackage{mathrsfs}
\usepackage{stackrel}
\usepackage{mathtools}
\usepackage{nicefrac}
\usepackage{thmtools}
\usepackage{stmaryrd}
\usepackage{wasysym}
\usepackage{pifont}
\usepackage{array}
\usepackage{cases}
\usepackage{adjustbox}
\usepackage[inline]{enumitem}
\usepackage{url}
\usepackage{doi}
\usepackage{acronym}
\usepackage{todonotes}

\usepackage{graphicx} 
\usepackage{hyperref} 
\usepackage[firstpage]{draftwatermark} 

\usepackage[font=small,labelfont=bf,tableposition=top]{caption}
\DeclareCaptionLabelFormat{andtable}{#1~#2  \&  \tablename~\thetable}

\usepackage{framed}
\usepackage{listings}
\usepackage{tabularx}
\usepackage{booktabs}
\usepackage{multirow}

\usepackage{tikz}
\usetikzlibrary{positioning}
\usetikzlibrary{arrows,automata}
\usetikzlibrary{shapes.geometric} 
\usetikzlibrary{shapes.misc} 
\usetikzlibrary{decorations.pathmorphing} 
\usetikzlibrary{matrix}
\usetikzlibrary{calc}

\usepackage{comment}

\usepackage{float}

\usepackage[frozencache, cachedir=.]{minted}

\usepackage{xcolor}

\newcommand{\slimparagraph}[1]{\noindent \emph{#1}}

\let\doteqrel\doteq
\renewcommand*{\doteq}{\mathbin{\doteqrel}}

\newcommand*{\lnextbase}{\ocircle}

\newcommand*{\lglobbase}{\square}
\newcommand*{\levenbase}{\Diamond}
\newcommand*{\lnextsup}[1]{\mathop{\lnextbase^{#1}}}

\newcommand*{\lguntil}[4]{#3 \mathbin{\mathcal{U}^{#1}_{#2}} #4}

\newcommand*{\lnext}{\mathop{\lnextbase}}

\newcommand*{\lluntil}[2]{#1 \mathbin{\mathcal{U}} #2}

\newcommand*{\llglob}{\mathop{\lglobbase}}

\newcommand*{\lleven}{\mathop{\levenbase}}

\newcommand*{\ldnext}{\lnextsup{d}}
\newcommand*{\lunext}{\lnextsup{u}}

\newcommand*{\lcnext}[1]{\mathop{\chi_F^{#1}}}
\newcommand*{\lcdnext}{\lcnext{d}}
\newcommand*{\lcunext}{\lcnext{u}}

\newcommand*{\lcduntil}[2]{{#1} \mathbin{\mathcal{U}_\chi^d} {#2}}

\newcommand*{\chain}{\chi}

\newcommand{\cmark}{\ding{51}}
\newcommand{\xmark}{\ding{55}}

\newcommand*{\pvar}[3]{\llbracket {#1}, {#2} \, | \, {#3} \rrbracket}
\newcommand*{\pvarnode}[2]{\llbracket {#1} \, | \, {#2} \rrbracket}
\newcommand*{\nex}[1]{\llbracket {#1} \, \mathord{\uparrow} \rrbracket}

\newcommand*{\lcall}{\mathbf{call}}
\newcommand*{\lret}{\mathbf{ret}}

\newcommand*{\lstm}{\mathbf{stm}}
\newcommand*{\lobs}{\mathbf{obs}}
\newcommand*{\lqry}{\mathbf{qry}}

\newcommand{\shortstackrel}[3][.3ex]{%
  \mathrel{\vbox{\offinterlineskip\ialign{%
    \hfil##\hfil\cr
    $\scriptscriptstyle{#2}$\cr
    \noalign{\kern#1}
    $#3$\cr
}}}}
\newcommand{\apush}[1]{\shortstackrel{#1}{\rightarrow}}
\newcommand{\ashift}[1]{\shortstackrel{#1}{\dashrightarrow}}
\newcommand{\apop}[1]{\shortstackrel{#1}{\Rightarrow}}
\newcommand{\asupp}[1]{\stackrel{#1}{\leadsto}}

\newcommand{\suppedge}{\rightarrow}

\newcommand{\symb}[1]{\mathop{smb}(#1)}

\newcommand{\toolname}{\texttt{POPACheck}}
\newcommand{\B}{\textcolor{blue}{\texttt{B}}\texttt{()}}
\newcommand{\arr}{\texttt{a}[\texttt{left..right}]}

\acrodef{ACP}{Algebra of Communicating Processes}
\acrodef{AP}{Atomic Proposition}
\acrodef{BDD}{Boolean Decision Diagram}
\acrodef{BPA}{Basic Process Algebra}
\acrodef{BSCC}{Bottom SCC}
\acrodef{CEGAR}{Counterexample-Guided Abstraction Refinement}
\acrodef{CFG}{Context-Free Grammar}
\acrodef{CFL}{Context-Free Language}
\acrodef{CPU}{Central Processing Unit}
\acrodef{CTL}{Computation Tree Logic}
\acrodef{DCFL}{De\-ter\-min\-is\-tic Context-Free Language}
\acrodef{DFS}{Depth-First Search}
\acrodef{DHP}{Downward Hierarchical Path}
\acrodef{DSP}{Downward Summary Path}
\acrodef{DS}{Downward Summary}
\acrodef{EF}{Ehrenfeucht-Fra\"iss\'e}
\acrodef{ERSM}{(Extended) Recursive State Machine}
\acrodef{ETR}{Existential first-order Theory of the Reals}
\acrodef{FOL}{First-Order Logic}
\acrodef{FO}{First-Order}
\acrodef{FSA}{Finite-State Automaton}
\acrodefplural{FSA}[FSA]{Finite-State Automata}
\acrodef{FSM}{Finite-State Machine}
\acrodef{JDK}{Java Development Kit}
\acrodef{lhs}{left-hand side}
\acrodef{LIFO}{Last In First Out}
\acrodef{LR}{Left-Recursive}
\acrodef{LTL}{Linear Temporal Logic}
\acrodef{MC}{Model Checking}
\acrodef{MSOL}{Monadic Second-Order Logic}
\acrodef{MSO}{Monadic Second-Order}
\acrodef{NBA}{Nondeterministic B\"uchi Automaton}
\acrodefplural{NBA}[NBA]{Nondeterministic B\"uchi Automata}
\acrodef{NWA}{Nested Words Automaton}
\acrodefplural{NWA}[NWA]{Nested Words Automata}
\acrodef{NWTL}{Nested Words Temporal Logic}
\acrodef{OPA}{Operator Precedence Automaton}
\acrodefplural{OPA}[OPA]{Operator Precedence Automata}
\acrodef{OPG}{Operator Precedence Grammar}
\acrodef{OPL}{Operator Precedence Language}
\acrodef{OPM}{Operator Precedence Matrix}
\acrodef{OPTL}{Operator Precedence Temporal Logic}
\acrodef{OP}{Operator Precedence}
\acrodef{OVI}{Optimistic Value Iteration}
\acrodef{PDA}{Pushdown Automaton}
\acrodefplural{PDA}[PDA]{Pushdown Automata}
\acrodef{PDS}{Pushdown System}
\acrodefplural{PDS}[PDS's]{Pushdown Systems}
\acrodef{pOPA}{Probabilistic Operator Precedence Automaton}
\acrodefplural{pOPA}[pOPA]{Probabilistic Operator Precedence Automata}
\acrodef{POMC}{Precedence Oriented Model Checker}
\acrodef{POTL}{Precedence Oriented Temporal Logic}
\acrodef{POTLF}[POTLf$\chain$]{Precedence Oriented Temporal Logic}
\acrodef{pPDA}{pro\-ba\-bi\-li\-stic Pushdown Automaton}
\acrodefplural{pPDA}[pPDA]{pro\-ba\-bi\-li\-stic Pushdown Automata}
\acrodef{PPL}{Probabilistic Programming Language}
\acrodef{PR}{Precedence Relation}
\acrodef{RAM}{Random-Access Memory}
\acrodef{rhs}{right-hand side}
\acrodefplural{rhs}[rhs's]{right-hand sides}
\acrodef{RMC}{Recursive Markov Chain}
\acrodef{RR}{Right-Recursive}
\acrodef{RSM}{Recursive State Machine}
\acrodef{SAT}{Satisfiability}
\acrodef{SCC}{Strongly Connected Component}
\acrodef{SMT}{Satisfiability Modulo Theories}
\acrodef{ST}{Syntax Tree}
\acrodef{TS}{Transition System}
\acrodef{UHP}{Upward Hierarchical Path}
\acrodef{UML}{Unified Modeling Language}
\acrodef{UOT}{Unranked Ordered Tree}
\acrodef{USP}{Upward Summary Path}
\acrodef{US}{Upward Summary}
\acrodef{VLTL}{Visibly Linear Temporal Logic}
\acrodef{VPA}{Visibly Pushdown Automaton}
\acrodefplural{VPA}[VPA]{Visibly Pushdown Automata}
\acrodef{VPL}{Visibly Pushdown Language}
\acrodef{OOPL}[$\omega$OPL]{Operator Precedence $\omega$-Language}
\acrodef{OPBA}[$\omega$OPBA]{Operator Precedence B\"uchi Automaton}
\acrodefplural{OPBA}[$\omega$OPBA]{Operator Precedence B\"uchi Automata}
\acrodef{OVPL}[$\omega$VPL]{Visibly Pushdown $\omega$-Language}
\acrodef{DRA}[DRA]{Deterministic Rabin Automaton}
\acrodefplural{DRA}[DRA]{Deterministic Rabin Automata}

\begin{document}
\title{\toolname{}: A Model Checker for Probabilistic Pushdown Automata}
%
%
\author{Francesco Pontiggia\inst{1}\orcidID{0000-0003-2569-6238} \and \\
    Ezio Bartocci\inst{1}\orcidID{0000-0002-8004-6601} \and \\
    Michele Chiari\inst{1}\orcidID{0000-0001-7742-9233}}
\institute{TU Wien, Vienna, Austria
\email{name.surname@tuwien.ac.at}}

\maketitle              

\acused{POTLF}

\SetWatermarkAngle{0}
\SetWatermarkText{\raisebox{15.5cm}{
\href{https://doi.org/10.5281/zenodo.15213850}{\includegraphics{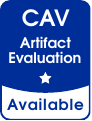}}
\hspace{7.5cm}
\includegraphics{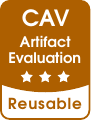}
}}

\begin{abstract}
We present \toolname{}, the first model checking tool for \acp{pPDA} supporting temporal logic specifications.
\toolname{} provides a user-friendly probabilistic modeling language with recursion that automatically translates into \acp{pOPA}.
\acp{pOPA} are a class of \acp{pPDA} that can express all the behaviors of probabilistic programs: sampling, conditioning, recursive procedures, and nested inference queries.
On \acp{pOPA}, \toolname{} can solve reachability queries as well as qualitative and quantitative model checking queries for specifications in \ac{LTL} and a fragment of \ac{POTL}, a logic for context-free properties such as pre/post-conditioning.

\end{abstract}

\section{Introduction}
\label{sec:intro}

\begin{wrapfigure}[10]{r}{0.4\textwidth}
\vspace{-10ex}
\begin{minted}[fontsize=\scriptsize]{python}
# global variables: a[], mid, val
B(u4 left, u4 right) {
  mid = Uniform(left, right);
  if (left < right){
    if (a[mid] < val) {
      left = min(mid +1, right);
      B(left, right);
    } else { if (a[mid] > val) {
      right = max(mid -1, left);
      B(left, right);
    } }
  }
}
\end{minted}
\vspace{-3ex}
\caption{Sherwood Binary search.}
\label{fig:Sherwood}
\vspace{-3em}
\end{wrapfigure}

The last two decades saw great efforts towards the analysis of \acfp{pPDA} ~\cite{EsparzaKM04,KuceraEM06,BrazdilEK05,BrazdilBHK08,BrazdilEKK13,BrazdilKKV15}, and the equivalent model of \acp{RMC}~\cite{EtessamiY05,YannakakisE05,EtessamiY05b,EtessamiY09,EtessamiY12,EtessamiY15} as a succinct formalism to express infinite-state probabilistic systems and to model probabilistic programs with nested and possibly recursive procedures. However, no existing tool implements the presented temporal logic model checking algorithms, due to their computational complexity and many practical obstacles. With \toolname{}, we leverage various recent results on \acp{pPDA} analysis to produce an efficient model checking tool that scales beyond toy examples, and forms a baseline for future verification tools for infinite-state probabilistic systems.

\slimparagraph{Illustrative Example.}
Consider the Sherwood~\cite{mcconnell2007analysis} variant of the binary search (Fig.~\ref{fig:Sherwood}), a well-known recursive algorithm.
\B{} searches for \texttt{val} in the array \arr{}.
Unlike the deterministic version, in each iteration \B{} selects the pivot \texttt{mid} randomly among the remaining portion of \texttt{a}.
Thanks to randomness, worst-, best- and average runtime align.
If \texttt{a} has a finite bound and its elements have a finite domain, \B{} can be translated into a finite-state \ac{pPDA}:
automaton states model global and local variable values,
while stack symbols model procedure parameter values.
A relevant property for this program is \textbf{partial correctness}~\cite{OlmedoKKM16}:
when \B{} is invoked in a state where \texttt{left} $\leq$ \texttt{right},
\arr{} is sorted and \texttt{val} occurs in \arr{}, then at return \texttt{mid} stores the index of \texttt{a} where \texttt{val} lies.
This property, just like pre/post-conditioning, requires to match each call to \B{} in the recursion
to its corresponding return, and skip inner calls to \B{} in the execution trace. It is a context-free property beyond \acs{LTL}'s expressivity, limited to regular properties.
\ac{POTL}~\cite{ChiariMP21a} is an expressive logic based on \acp{OPL}~\cite{Floyd1963}, a subclass of context-free languages, and expresses partial correctness as:
\begin{center}
$\llglob (\lcall \land \mathtt{B} \land \mathtt{sorted} \land \mathtt{valOccurs} \land  \mathtt{left} \leq \mathtt{right} \implies \lcunext \mathtt{a[mid]} == \mathtt{val})$
\end{center}

\slimparagraph{Probabilistic programming.}
Probabilistic programs have recently gained popularity in AI and machine learning~\cite{Ghahramani15}, where they are employed as succinct models for Bayesian inference~\cite{GordonHNR14}.
In addition to randomized assignment, a prominent feature of this programming model is \emph{conditioning},
which allows for adding evidence of observed events by conditioning the program variables to take certain values.
When using rejection sampling, a probabilistic program contains statements \texttt{observe(e)},
where \texttt{e} is a Boolean condition.
If \texttt{e} is not satisfied, the current execution trace is rejected: in such a case, we assume that the program is restarted.
The semantics of a probabilistic program is the probability distribution in the return statement, also called \emph{posterior}.
Probabilistic programs can also be nested: a program samples a value from a distribution represented by another probabilistic program.
They are known under the name of \emph{nested queries} \cite{GoodmanMRBT08}, and model scenarios beyond the expressivity of flat probabilistic programs: most notably, metareasoning patterns (the so called Theory of Mind)~\cite{StuhlmullerG14,ZhangA22}, linguistics~\cite{speechact}, cognitive science~\cite{probmods2}, multi-agent planning~\cite{seaman2020} and sequential decision making~\cite{Evans17}.
To the best of our knowledge, \acp{pOPA}~\cite{abs-2404-03515} are the smallest subclass of \acp{pPDA} that can model effectively these behaviors. \acp{pOPA} traces are \acp{OPL}.
Unlike general context-free languages, \acp{OPL} are closed under most relevant Boolean operations~\cite{MP18}.
Let $\mathcal{B}$ be an (deterministic or separated) \ac{OPBA}, the class of automata recognizing \acp{OPL}.
When a \acp{pOPA} $\mathcal{A}$ is defined over the same alphabet as $\mathcal{B}$, we can verify automatically $\mathcal{A}$ against $\mathcal{B}$ via automata-based model checking~\cite{abs-2404-03515}.

\slimparagraph{Our approach.} The main challenge in \acp{pPDA} model checking is computing the \emph{termination probabilities}, which involves solving a system of \textbf{non-linear} equations. 
While tools like PReMo~\cite{WojtczakE07} approximate them from below, for general model checking we need to know whether these quantities sum up to exactly 1.
We could employ off-the-shelf SMT solvers (e.g. Z3~\cite{MouraB08}) with decision procedures for the \ac{ETR} (\textsc{qf\_nra}), but they offer only doubly exponential decision procedures~\cite{JovanovicM12},
although \ac{ETR} is in \textsc{pspace}~\cite{Canny88,Renegar92}.
We thus devise a semi-algorithm leveraging both certificates for termination probabilities~\cite{WinklerK23a} via a numeric method called \ac{OVI}, and certificates for expected runtimes~\cite{WinklerK23b}.
A second challenge is the need for deterministic automata (or weak variants thereof) in probabilistic verification---a common problem already occurring in the far simpler setting of \acs{LTL} model checking for Markov Chains \cite{CouvreurSS03,BaierK00023,BaierK08}.
In this regard, we exploit~\cite{EtessamiY12,abs-2404-03515} which provide resp.\ single exponential model checking algorithms (i.e., avoiding determinization) for \acs{LTL} and a fragment of \ac{POTL} (\acs{POTLF}).
These algorithms represent the theoretical ground of \toolname{}.

\toolname{} is hosted on Github at \url{https://github.com/michiari/POMC/}. It is an extension of the POMC tool~\cite{PontiggiaCP21,Pontiggia21,ChiariMPP23}, and relies on its modules for constructing automata from formulae.

\slimparagraph{Contributions.}
We present i) a semi-algorithm overcoming numerical issues regarding computing termination probabilities;
ii) \toolname, a model checker for \acp{pPDA} based on this semi-algorithm; 
iii) a user-friendly domain-specific language for recursive probabilistic programs;
and iv) an extensive experimental evaluation with a benchmark of programs and LTL/\acs{POTLF} formulae.

\section{Background}
\label{sec:background}

\subsection{\acfp{pOPA}}
\acp{pOPA} are \acp{pPDA} with state labels from a set $\Sigma$.
State labels drive the stack behavior of \acp{pOPA} through three \emph{\acp{PR}}:
given two labels $a$ and $b$, we say $a$ \emph{yields precedence} to $b$ iff $a \lessdot b$,
$a$ and $b$ are \emph{equal in precedence} iff $a \doteq b$,
and $a$ \emph{takes precedence} from $b$ iff $a \gtrdot b$.
An \ac{OPM} is a total function
$M : \Sigma^2 \rightarrow \{\mathord{\lessdot}, \mathord{\doteq}, \mathord{\gtrdot}\}$;
the special symbol $\#$ is s.t.\ $\# \lessdot a$ for all $a \in \Sigma$.
In the following, let $\mathfrak{D}(S) = \{f : S \rightarrow [0,1] \mid \sum_{s \in S} f(s) = 1\}$ denote he set of probability distributions on a finite set $S$.
\begin{definition}[\cite{abs-2404-03515}]
A \ac{pOPA} is a tuple
$\mathcal A = (\Sigma, \allowbreak M, \allowbreak Q, \allowbreak u_0,
\allowbreak \delta, \allowbreak \Lambda)$ where:
$\Sigma$ is a finite set of state labels; $M$ is an \ac{OPM};
$Q$ is a finite set of states;
$u_0$ is the initial state;
$\Lambda : Q \rightarrow \Sigma$ is a state labeling function; and
$\delta$ is a triple of transition functions
$\delta_{\mathit{push}} : Q \rightarrow \mathfrak{D}(Q)$,
$\delta_{\mathit{shift}} : Q \rightarrow \mathfrak{D}(Q)$, and
$\delta_{\mathit{pop}} : (Q \times Q) \rightarrow \mathfrak{D}(Q)$,
such that pop moves have the following condition, for all $u, s, v \in Q$:
\vspace{-1ex}
\begin{center}
\(
\delta_\mathit{pop}(u, s)(v) > 0 \implies
  \forall a \in \Sigma : a \gtrdot \Lambda(u) \implies a \gtrdot \Lambda(v).
\)
\vspace{-.5ex}
\end{center}
\end{definition}
The semantics of $\mathcal{A}$ is
an infinite Markov chain~\cite{BaierK08} $\Delta(\mathcal{A})$
with vertex set $Q \times (\Gamma^* \{\bot\})$
where $\bot$ is the initial stack symbol,
which can never be pushed or popped,
and $\Gamma = \Sigma \times Q$ is the set of stack symbols.
\acp{PR} decide whether to push onto the stack, update the topmost symbol, or pop from it.
For any stack contents $A \in \Gamma^* \{\bot\}$:

\smallskip
\noindent\textbf{push:} $(u, A) \apush{x} (v, [\Lambda(u), u] A)$
  if $\symb{A} \lessdot \Lambda(u)$
  and $\delta_{\mathit{push}}(u)(v) = x$;\\
\textbf{shift:} $(u, [a, s] A) \apush{x} (v, [\Lambda(u), s] A)$
  if $a \doteq \Lambda(u)$
  and $\delta_{\mathit{shift}}(u)(v) = x$;\\
\textbf{pop:} $(u, [a, s] A) \apush{x} (v, A)$
  if $a \gtrdot \Lambda(u)$
  and $\delta_{\mathit{pop}}(u, s)(v) = x$;

\smallskip
\noindent where $\symb{\bot} = \#$ and $\symb{[a, r]} = a$ for $[a, r] \in \Gamma$.

Stack symbols are pairs of a label and a state,
and the \ac{PR} between the label in the topmost stack symbol
and that of the current state decides the next move.
If such \ac{PR} is $\lessdot$, a push move puts the current state and its label onto the stack.
Since $\symb{\bot} = \#$, the first move is always a push.
If the \ac{PR} is $\doteq$, a shift move updates the topmost stack symbol by replacing its label with the current state's.
Finally, if the \ac{PR} is $\gtrdot$, the topmost stack symbol gets popped.

A run of $\mathcal{A}$ is a path in $\Delta(\mathcal{A})$ starting in $(u_0, \bot)$.
The probability space on the set of runs
is obtained by the cylinder set construction as for Markov chains~\cite{BaierK08}.
The set of infinite words formed by labels of states in a run is an \ac{OPL}~\cite{abs-2404-03515}.

\begin{wrapfigure}[7]{r}{3.5cm}%
\vspace{-4ex}
\scriptsize
\centering
\(
\begin{array}{r | c c c c c}
         & \lcall   & \lret   & \lqry    & \lobs    & \lstm \\
\hline
\lcall   & \lessdot & \doteq  & \lessdot & \gtrdot  & \lessdot \\
\lret    & \gtrdot  & \gtrdot & \gtrdot  & \gtrdot  & \gtrdot \\
\lqry    & \lessdot & \doteq  & \lessdot & \lessdot & \lessdot \\
\lobs    & \gtrdot  & \gtrdot & \gtrdot  & \gtrdot  & \gtrdot \\
\lstm    & \gtrdot  & \gtrdot & \gtrdot  & \gtrdot & \gtrdot \\
\end{array}
\)
\caption{\acs{OPM} $M_\lcall$.}
\label{fig:opm}
\end{wrapfigure}
\acp{PR} guide the stack behavior of \acp{pOPA},
and \acp{PR} between state labels completely determine whether the \ac{pOPA} pushes, updates, or pops a stack symbol.
This dependency of the stack behavior on labels, and hence traces, enables the definition of stack-aware context-free modalities in the specification formalism (\acs{POTL}),
by allowing the definition of a synchronized product between \ac{pOPA} and a pushdown automaton encoding the specification (cf.~\cite{abs-2404-03515}).
We thus define state labels to describe events that affect the program stack,
and we define \acp{PR} between state labels so that such events have the same effect on the \ac{pOPA} stack that they have on the stack of activation frames in a high-level procedural probabilistic programming language with rejection sampling.
\footnote{Note that \acp{PR} are not ordering relations, so they need not enjoy standard properties such as reflexivity and transitivity.}

State labels in \ac{OPM} $M_\lcall$ (Fig.~\ref{fig:opm}) represent traces of probabilistic programs.
$\lcall$ and $\lret$ are function calls and returns,
$\lstm$ statements that do not affect the stack (e.g., randomized assignments).
$\lqry$ is the \emph{conditional sampling} operator (roughly corresponding to \texttt{query} in Church \cite{GoodmanMRBT08}): it reifyies the posterior distribution of a probabilistic program, allowing for sampling from it. $\lobs$ denotes an unsatisfied observation.
The table is to be read by row---e.g., we have $\lcall \doteq \lret$ and $\lret \gtrdot \lcall$.
\acp{PR} are such that a \ac{pOPA} always pushes from a $\lcall$ state
($\lcall \lessdot \lcall$, etc.),
and pops after being in a $\lret$ state through a shift move
($\lcall \doteq \lret$ and $\lret$ takes precedence from other symbols).
This way, the \ac{pOPA} stack mimics the program's stack.
$\lstm$ causes a push immediately followed by a pop (because $\lstm \gtrdot$ all labels),
thus leaving the stack unchanged.
Moreover, $\lobs$ triggers pop moves that unwind the stack until they reach the first $\lqry$ symbol.
Thereby, the \acs{pOPA} excludes (or rejects) the trace at hand from the posterior distribution of the current probabilistic program without affecting outer queries, effectively simulating nested rejection sampling~\cite{OlmedoGJKKM18}.

\subsection{Specification Formalism}

We express specifications through a temporal logic with the following syntax:
\begin{center}
\vspace{-1ex}
\(
    \varphi \coloneqq \mathrm{a}
    \mid T
    \mid \neg \varphi
    \mid \varphi \lor \varphi
    \mid \lnext \varphi
    \mid \lluntil{\varphi}{\varphi}
    \mid \lnextsup{t} \varphi
    \mid \lcnext{t} \varphi
    \mid \lguntil{t}{\chi}{\varphi}{\varphi}
\)
\vspace{-1ex}
\end{center}
Formulae are evaluated in the first position of program traces,
and subformulae in further positions.
$\mathrm{a}$ is an \emph{atomic proposition} from a finite set $AP$ containing
labels from Fig.~\ref{fig:opm} and names of program functions.
$T$ is any term of the form $e_1 \bowtie e_2$, where $\bowtie$ is a binary comparison operator
($=$, $>$, etc.) and $e_1, e_2$ are integer arithmetic expressions involving program variables and constants.
$\neg$ and $\lor$ are propositional operators,
to which we add the derived operators $\land$ and $\implies$ with the usual semantics.
$\lnext \varphi$ and $\lluntil{\varphi_1}{\varphi_2}$ are the \emph{next} and \emph{until} operators from \ac{LTL}~\cite{Pnueli77},
resp.\ meaning that $\varphi$ will hold in the next time instant,
and that $\varphi_1$ holds until a time instant in which $\varphi_2$ does.
We use the derived operators \emph{eventually} $\lleven \varphi \equiv \lluntil{\top}{\varphi}$,
and \emph{globally} $\llglob \varphi \equiv \neg \lleven \neg \varphi$.

The remaining operators form a fragment of \ac{POTL} called \acs{POTLF}.
They move up ($t = u$) and down ($t = d$) among function frames in the program stack.
$\ldnext \varphi$ holds if the next time instant belongs to the same or a lower frame
and $\varphi$ holds in it, and \emph{vice versa} for $\lunext$.
$\lcdnext$ moves along a binary relation that links function calls to their returns and inner function calls.
\begin{wrapfigure}[6]{r}{0.3\textwidth}
\vspace{-7ex}
\begin{minted}[fontsize=\scriptsize]{c}
f1() {
     ... // other statements
     f2(x);
     ... 
     return y;
}
\end{minted}
\vspace{-3ex}
\caption{Program stub.}
\label{fig:chain-prog}
\vspace{-3em}
\end{wrapfigure}
In the program of Fig.~\ref{fig:chain-prog},
the relation links events representing $\lcall$s to \texttt{f1()}
to those representing the inner $\lcall$ to \texttt{f2()} and \texttt{f1}'s return statement.
E.g., if evaluated in an instant labeled with a $\lcall$ to \texttt{f1},
$\lcdnext (\lcall \land \mathtt{x} > 0)$ holds if $\mathtt{x} > 0$ in the instant when \texttt{f1} calls \texttt{f2},
and $\lcdnext (\lret \land \mathtt{y} > 0)$ holds if \texttt{f1} returns in an instant in which $\mathtt{y} > 0$.
Moreover, $\lcunext \lobs$ holds if a false \texttt{observe} is encountered
anytime during the execution of \texttt{f1}.
The semantics of $\lguntil{t}{\chi}{}{}$ are the same as the \ac{LTL} until,
except it works on paths obtained by iterating the $\lnextsup{t}$ and $\lcnext{t}$ operators,
for $t \in \{u,d\}$.
For a complete definition, cf.~\cite{ChiariMP21b}.

\section{Model Checking algorithm}
\label{sec:mc-algorithm}

\subsection{The Support Chain}
\label{sec:support-chain}

Let $\mathcal{A} = (\Sigma, \allowbreak M, \allowbreak Q, \allowbreak u_0,
\allowbreak \delta, \allowbreak \Lambda)$
be a \ac{pOPA}.
A \emph{support} is a sequence of \ac{pOPA} moves
\(
u_0
\apush{}{u_1}
\ashift{}{}
\dots
\ashift{}{u_\ell}
\apop{u_0} {u_{\ell+1}},
\)
denoted as $u_0 \asupp{}{} u_{\ell+1}$,
that occur from the move pushing a stack symbol $[\Lambda(u_0), u_0]$ to the move popping it.
Given $u, v \in Q$ and $\alpha \in \Gamma$, we define $\pvar{u}{\alpha}{v}$
as the probability that $\mathcal{A}$ starting in $(u, \alpha \bot)$ reaches $v$ at the end of the support that put $\alpha$ on top of the stack.
Such \emph{termination probabilities}~\cite{EtessamiY09,BrazdilEKK13}
are the least non-negative solutions of the equation system
$\mathbf{v} = f(\mathbf{v})$, where $\mathbf{v}$ is the vector of triples
$\pvar{u}{\alpha}{v}$ for all $u, v \in Q$, $\alpha \in \Gamma$ and $f(\pvar{u}{\alpha}{v})$ is given by
\vspace{-1ex}
\[
\begin{cases}
\sum_{r, t \in Q} \delta_\mathit{push}(u)(r) \pvar{r}{[\Lambda(u), u]}{t} \pvar{t}{\alpha}{v} & \text{if $\alpha = \bot$, or $\alpha = [a, s]$ and $a \lessdot \Lambda(u)$} \\
\sum_{r \in Q} \delta_\mathit{shift}(u)(r) \pvar{r}{[\Lambda(u), s]}{v} & \text{if $\alpha = [a, s]$ and $a \doteq \Lambda(u)$} \\
\delta_\mathit{pop}(u, s)(v) & \text{if $\alpha = [a, s]$ and $a \gtrdot \Lambda(u)$}
\end{cases}
\vspace{-1ex}
\]
The probability that $\alpha \in \Gamma$ is never popped
after a run visits a configuration $(u, \alpha A)$ for $u \in Q$ and some stack contents $A$ is
\(\nex{u, \alpha} = 1 - \sum_{v \in Q} \pvar{u}{\alpha}{v}.\)

The \emph{support chain} is a finite Markov chain that replaces supports with single transitions,
and describes the behavior of 
\ac{pOPA} runs while preserving their probability distribution.
\begin{definition}[\cite{abs-2404-03515}]
The \emph{support chain} $M_\mathcal{A}$ of $\mathcal{A}$ is a Markov chain with states in
$\mathcal{C} = \{ (u, \alpha) \in Q \times \Gamma_\bot \mid \nex{u, \alpha} > 0 \}$,
initial state $(u_0, \bot)$,
and a transition probability function $\delta_{M_\mathcal{A}}$ such that
\begin{itemize}
\item
  $\delta_{M_\mathcal{A}}(u, [a, s])(v, [\Lambda(u), s]) = \delta_{\mathit{shift}}(u)(v) \nex{v, [\Lambda(u), s]} / \nex{u, [a, s]}$\\
  for all $(u, [a, s]), (v, [\Lambda(u), s]) \in \mathcal{C}$ such that $a \doteq \Lambda(u)$;
\item otherwise, for all $(u, \alpha), (v, \alpha') \in \mathcal{C}$,\\
  \(
  \delta_{M_\mathcal{A}}(u, \alpha)(v, \alpha') =
    (P_\mathit{push} + P_\mathit{supp}) \nex{v, \alpha'} / \nex{u, \alpha}
  \)
  where
  \begin{itemize}
    \item $P_\mathit{push} = \delta_{\mathit{push}}(u)(v)$ if $\symb{\alpha} \lessdot \Lambda(u)$,
          and $P_\mathit{push} = 0$ otherwise;
    \item $P_\mathit{supp} = \sum_{v' \in Q} \delta_{\mathit{push}}(u)(v') \pvar{v'}{[\Lambda(u), u]}{v}$,
          if $\alpha = \alpha'$, $\symb{\alpha} \lessdot \Lambda(u)$, and $\mathcal{A}$ has a support $u \asupp{}{} v$,
          and $P_\mathit{supp} = 0$ otherwise.
  \end{itemize}
\end{itemize}
\end{definition}

\begin{figure}[tb]
\centering
\begin{tikzpicture}
  [ node distance=0pt, font=\scriptsize, >=latex,
    action/.style={draw, rectangle},
    cond/.style={draw, chamfered rectangle, chamfered rectangle xsep=5em, chamfered rectangle ysep=0pt},
    scale=.7
  ]
\node (popa) [action] {pOPA};
\node (fv) [action, right=10pt of popa] {$f(\mathbf{v})$};
\node (Z3) [action, right=11pt of fv, yshift=10pt] {Z3 $\overline{\mathbf{v}}^*$};
\node (ovi) [action, right=10pt of fv, yshift=-10pt] {\textsc{ovi} $\overline{\mathbf{v}}^*$};
\node (ast) [cond, right=10pt of ovi, yshift=10pt] {$\underline{\nex{c}} > 0?$};
\node (past) [action, right=10pt of ast] {Z3 (\textsc{past})};
\node (pastc) [cond, right=10pt of past] {is $c$ \textsc{past}?};
\node (inc) [action, right=10pt of pastc, yshift=-10pt, text width=42pt] {inconclusive};
\node (disc) [action, above=3pt of inc, text width=42pt] {discard};
\node (suppc) [action, above=3pt of disc, text width=42pt] {add to SC};
\path (popa) edge[->] (fv)
      (fv) edge[->] (Z3)
      (Z3) edge[->] (ast.west)
      (fv) edge[->] (ovi)
      (ovi) edge[->] (ast.west)
      (ast) edge[->] node[shift={(-1pt,-3pt)}]{no} (past)
      (past) edge[->] (pastc);
\path[draw, ->] (ast.east) node[shift={(6pt,16pt)}]{yes} |- (suppc.west);
\path[draw, ->] (pastc.east) node[shift={(3pt,8pt)}]{yes} |- (disc.west);
\path[draw, ->] (pastc.east) node[shift={(5pt,-7pt)}]{no} |- (inc.west);
\draw[dotted] ($ (Z3.north west) + (-3pt,15pt) $) node[shift={(12pt,-4pt)}]{$\forall c \in \mathcal{C}$} rectangle ($ (inc.south east) + (1pt,-1pt) $);
\end{tikzpicture}
\caption{Overview of the semi-algorithm for building the support chain (SC).}
\label{fig:workflow-diagram}
\end{figure}
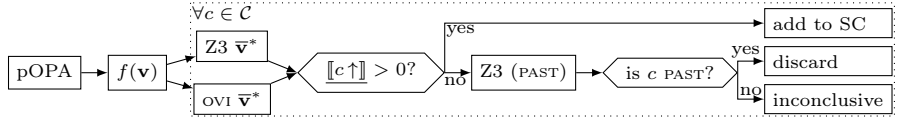

During infinite runs of $\mathcal{A}$, the stack contains symbols that are never popped.
States of $M_\mathcal{A}$ consist of such symbols,
and the current state of $\mathcal{A}$ when they are put on the stack.
$M_\mathcal{A}$ has three types of transitions:
push and shift moves that put on the stack (resp.\ update) a symbol that will never be popped,
and \emph{support} edges, which summarize the finite portion of a run
from when a stack symbol is pushed to when it is popped---a \emph{support}.
Probabilities assigned to edges $c \suppedge{} c'$ are conditioned on the fact that
the stack symbol in the target node $c'$ is never popped,
given that the one in the starting node $c$ is also never popped:
hence all probabilities are multiplied by $\nex{c'}/\nex{c}$.
Probabilities for push and shift edges follow directly from $\mathcal{A}$'s $\delta$ distributions,
while those for support edges include $P_\mathit{supp}$,
the probability that the stack symbol underlying the edge is popped.
Thus, $M_\mathcal{A}$ describes the limit behavior of $\mathcal{A}$ by only considering permanent stack symbols and transitions that created them, while summarizing with single edges run portions that involve stack symbols that are eventually popped.

To build the support chain of $\mathcal{A}$, 
we must check whether $\nex{c} > 0$ for all $c \in Q \times \Gamma_\bot$.
We propose the approach sketched in Fig.~\ref{fig:workflow-diagram}.
We further optimize this algorithm by applying all steps bottom-up to \acp{SCC} of the graph of nodes $c$.


Let $\mathbf{v}^*$ be the \emph{least} solution of $\mathbf{v} = f(\mathbf{v})$,
i.e., the vector of termination probabilities $\pvarnode{c}{v}$.
Recall that $\nex{c} = 1 - \sum_{v \in Q} \pvarnode{c}{v}$.
First, we find an upper bound $\overline{\mathbf{v}}^*$ for $\mathbf{v}^*$,
which we use to compute a lower bound $\underline{\nex{c}}$ for $\nex{c}$.
If $\underline{\nex{c}} > 0$, then $\nex{c} > 0$ and we add $c$ to the support chain.
Otherwise, we make the hypothesis that $\nex{c} = 0$,
i.e., that $c$ is almost-surely terminating (\textsc{ast}),
and check the stronger condition that $c$ is positively \textsc{ast} (\textsc{past}). 
A node $c = (u, \alpha)$ is \textsc{past} if stack symbol $\alpha$ is popped with probability 1 in \emph{finite expected time}.
We check \textsc{past} by solving with Z3 a system of linear equations~\cite{WinklerK23b} for the expected termination times.
If a solution exists, then $c$ is \textsc{past} and therefore \textsc{ast}: 
we do not add $c$ to the support chain.
Otherwise, the whole algorithm is inconclusive.

We find the upper bound $\overline{\mathbf{v}}^*$ in two different ways.
In both cases, we first find a lower bound $\underline{\mathbf{v}}^*$ to $\mathbf{v}^*$ via a decomposed variant of Newton's method \cite{EtessamiY09,EsparzaKL10}, 
after cleaning the system from zero-solution variables via value iteration.
\begin{itemize}
\item In the first method, we give $\underline{\mathbf{v}}^*$ as a hint to Z3, and
obtain  $\overline{\mathbf{v}}^*$ as a model for $\mathbf{v}$ in the query
$\underline{\mathbf{v}}^* \leq \mathbf{v} \leq \underline{\mathbf{v}}^* + \varepsilon \land \mathbf{v} \geq f(\mathbf{v})$,
for a small positive $\varepsilon$.
\item The other method employs \ac{OVI}~\cite{WinklerK23a}, which computes $\overline{\mathbf{v}}^*$ numerically.
We run \ac{OVI} initialized on $\underline{\mathbf{v}}^*$.
\end{itemize}

\subsection{Avoid determinization with a separation-based approach}
\label{sec:mc-algorithm-separation}

A prominent approach to \acs{LTL} model checking on probabilistic systems expresses specifications as deterministic automata.
Storm~\cite{HenselJKQV22} and PRISM~\cite{KwiatkowskaNP11} translate the input formula into a \ac{DRA}~\cite{BaierK08}, 
with a worst-case doubly exponential blowup.
While singly exponential algorithms that avoid determinization exist~\cite{CouvreurSS03,BaierK00023},
a common argument for using the conceptually simpler notion of Rabin automata is the existence in practice of effective translations from LTL formulae to relatively small-sized \acp{DRA}~\cite{EsparzaKS20}.
Unfortunately, the same cannot be said for \ac{POTL} formulae: the automata construction for \acs{POTL} is much more involved~\cite{ChiariMPP23},
and easily generates intractable automata.

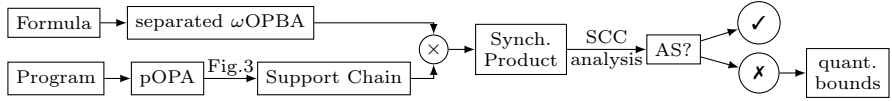
\begin{figure}[tb]
\centering
\begin{tikzpicture}
  [ node distance=0pt, font=\scriptsize, >=latex,
    action/.style={draw, rectangle},
    cond/.style={draw, chamfered rectangle, chamfered rectangle xsep=5em, chamfered rectangle ysep=0pt},
    round/.style={draw, circle},
    scale=.7
  ]
\node (phi) [action, align=center] {Formula};
\node (opba) [action, right=10pt of phi, align=center] {separated \acs{OPBA}};
\node (prog) [action, below=9pt of phi.south west, anchor=north west, align=center] {Program};
\node (popa) [action, right=10pt of prog, align=center] {\acs{pOPA}};
\node (sc) [action, right=20pt of popa, align=center] {Support Chain};
\node (prod) [round, inner sep=1pt, right=3pt of sc, yshift=11pt] {$\times$};
\node (g) [action, right=10pt of prod, align=center] {Synch.\\Product};
\node (ast) [action, right=30pt of g, align=center] {AS?};
\node (yes) [round, right=160pt of opba, align=center] {\cmark};
\node (no) [round, below=3pt of yes, align=center] {\xmark};
\node (quant) [action, right=10pt of no, align=center] {quant. \\ bounds};
\path (phi) edge[->] (opba)
      (prog) edge[->] (popa)
      (popa) edge[->] node[yshift=4pt] {Fig.\ref{fig:workflow-diagram}} (sc);
\path[draw, ->] (opba.east) -| (prod.north);
\path[draw, ->] (sc.east) -| (prod.south);
\path (prod) edge[->] (g)
      (g) edge[->] node[align=center] {\acs{SCC}\\analysis} (ast);
\path (ast) edge[->] (yes)
      (ast) edge[->] (no)
      (no) edge[->] (quant);
\end{tikzpicture}
\caption{Overview of model checking probabilistic programs against \acs{POTLF} formulae.}
\label{fig:qualitative-workflow-diagram}
\end{figure}

\toolname{} takes a different approach, outlined in Fig.~\ref{fig:qualitative-workflow-diagram}.
\acs{LTL} formulae can be translated into \emph{separated} B\"uchi automata,
i.e., such that the languages they accept starting from different states are disjoint~\cite{CouvreurSS03}.
Similarly, \acs{POTLF} formulae can be translated into \acp{OPBA},
i.e., pushdown automata that capture the class of infinite-word \acp{OPL},
that are \emph{separated}.
Thanks to separation, we can build a synchronized product
between the support chain of the \ac{pOPA} modeling the program
and the automaton encoding the formula.
We then perform qualitative model checking by analyzing the \acp{SCC} of the product.
Since formulae are translated into automata of exponential size~\cite{VardiW94,ChiariMPP23},
the whole procedure requires time exponential in formula size
and space polynomial in \ac{pOPA} size.
For quantitative model checking, we compute numerically the probabilities associated with edges
in the synchronized product:
edges subsuming supports require solving equation systems resembling those for termination probabilities.
We compute bounds for them with \ac{OVI}.
Since the product has size exponential in formula
length, quantitative model checking is in \textsc{expspace}.
Cf.~\cite{abs-2404-03515} for details.

\section{Input Language}
\label{sec:toolDescr}
\toolname{} analyzes programs written in MiniProb, a simple probabilistic programming language (Fig.~\ref{fig:miniprob}).
MiniProb supports (un)signed integer variables of arbitrary width
(\texttt{u8} is an 8-bit unsigned type) and fixed-size arrays.
Functions take parameters by value or value-result (with \&).
Actual parameters can only be variable identifiers for value-result parameters,
and any expression if passed by value.
Expressions consist of variables, array indexing, integer constants,
\begin{wrapfigure}[19]{r}{.5\textwidth}
\begin{minipage}{.45\textwidth}
\vspace{-7ex}
\small
\begin{align*}
\mathit{prog} \coloneqq & \; [\mathit{decl} ; \dots] \; \mathit{func} \; [\mathit{func} \dots] \\
\mathit{decl} \coloneqq & \; \mathit{type \; identifier} \; [, \mathit{identifier} \dots] \\
\mathit{type} \coloneqq & \; \mathtt{bool} \mid \mathtt{u}\mathit{int} \mid \mathtt{s}\mathit{int} \mid \mathtt{u}\mathit{int}[\mathit{int}] \mid \mathtt{s}\mathit{int}[\mathit{int}] \\
\mathit{func} \coloneqq & \; f \texttt{(}\mathit{type} \; [\&] x_1 \; [, \mathit{type} \; [\&] x_2 \dots]\texttt{)} \\
  &\; \{ [\mathit{decl} ; \dots] \; \mathit{block} \} \\
\mathit{stmt} \coloneqq
    & \; \mathit{lval} = e \\
    &\mid \mathit{lval} = \mathtt{Distribution}\texttt{(} \dots \texttt{)} \\
    &\mid \mathit{lval} = e_1 \{ e_2 : e_3 \} [e_4 \{ e_5 : e_6 \} \dots] e_n \\
    &\mid [\texttt{query}] \; f\texttt{(}e_1 \mid \mathit{lval}_1 \; [, e_2 \mid \mathit{lval}_2 \dots]\texttt{)} \\
    &\mid \mathtt{if} \; \texttt{(}e\texttt{)} \; \{ \mathit{block} \} \ \mathtt{else} \ \{ \mathit{block} \} \\
    &\mid \mathtt{while} \; \texttt{(}e\texttt{)} \; \{ \mathit{block} \} \\
    &\mid \mathtt{observe} \; \texttt{(}e\texttt{)} \\
\mathit{block} \coloneqq & \; \mathit{stmt} ; [\mathit{stmt} \dots ; ] \\
\mathit{lval} \coloneqq & \; \mathit{identifier} \mid \mathit{identifier}[e]
\end{align*}
\end{minipage}
\caption{MiniProb syntax.}
\label{fig:miniprob}
\end{wrapfigure}
and the usual arithmetic and Boo\-lean operators, including comparisons.
Boo\-lean operators handle integers (0 means false, everything else true).
Programs may sample from $\texttt{Bernoulli(} e_1, e_2 \texttt{)}$,
which returns 1 with probability $p = e_1 / e_2$, and 0 with probability $1-p$,
or from $\texttt{Uniform(} e_1, e_2 \texttt{)}$,
which samples uniformly among integers from $e_1$ to $e_2 - 1$.
Random assignments of the form $x = e_1 \{ e_2 / e_3 \} e_4$ mean that $x$ is assigned
the value of $e_1$ with probability $e_2 / e_3$, and $e_4$ with probability $1 - e_2 / e_3$.
Finally, functions can \texttt{query} the distribution on value-result parameters of another function,
and condition on a Boolean expression with \texttt{observe}.

\section{Evaluation}
\label{sec:experiments}
We devised our experiments around the following questions:
can our approach build support chains for probabilistic programs of medium size?
How large equations systems arise in quantitative model checking? 
How scalable is our implementation, i.e., how large programs can our tool solve in a reasonable time? 

\begin{table}[bt]
\caption{Benchmark formulae.}
\label{tab:formulae_experiments}
    \centering
    \begin{adjustbox}{max width=\textwidth}
    \renewcommand{\arraystretch}{1.1}
    \begin{tabular}{l @{\hspace{.4em}} l l @{\hspace{.4em}} l}
    \toprule
    \# & Formula & \# & Formula  \\
    \midrule
    \textbf{Q.1}  & $ \lleven \llglob (\neg \lobs)$ &
    \textbf{Q.2}  & $\llglob \bigl(\lqry \implies \ldnext (\lcall \land \neg \lcunext \lobs) \lor \lcdnext (\lcall \land \neg \lcunext \lobs)\bigr)$ \\
    \textbf{Q.3}  & $\llglob \, (\lcall \land \mathtt{Alice} \land \mathtt{p} \geq 0.4 \implies \neg \lcunext \lobs)$ & 
    \textbf{Q.4} & $\lluntil{\neg \mathtt{ sampleA }} {(\lcall \land \mathtt{sampleA} \land \lcdnext (\lqry \land \ldnext (\lcall \land \mathtt{sampleA})))}$ \\
    \textbf{Q.5} & $\neg \lleven \bigl(\lqry \land \lcunext (\mathtt{sampleA} \land \mathtt{opR} == 0 \land \mathtt{opC} == 1)\bigr)$ &
     \textbf{Q.6} & $\lcduntil{(\neg \mathtt{elder})}{(\mathtt{young} \land f)}$  \\
    \textbf{Q.7} & $\lleven \bigl(\lcdnext (\ldnext \mathtt{elder}) \land \lcunext (\mathtt{elder} \land \neg f)\bigr)$ &
    \textbf{Q.8} & $ \lcunext (\mathtt{aliceLoc} == 1)$ \\
    \textbf{Q.9} & $\lleven \bigl(\lret \land \mathtt{main} \land \mathtt{aliceLoc} == 1\bigr)$ &
    \textbf{Q.10} & $\neg \lcunext (\mathtt{R} == 0 \land \mathtt{C} == 2)$ \\
    \textbf{Q.11} & $\neg \lcunext (\mathtt{R} == 0 \land \mathtt{C} == 1)$ & \textbf{Q.12} & $\lluntil{\neg \lobs}{(\lcall \land \mathtt{Bob} \land \lcunext \lobs)}$ \\
     \textbf{Q.13} & $\lluntil{\neg \mathtt{Alice}}{(\lcall \land \mathtt{Alice} \land \neg \lcunext \lobs)}$ & \textbf{Q.14} & $\lleven (\lret \land \mathtt{main} \land \mathtt{R} == 0 \land \mathtt{C} == 2)$
     \\
    \textbf{Q.15} & $\lluntil{\neg \lobs}{(\lcall \land \mathtt{Alice} \land \lcunext \lobs)}$ && \\
    \textbf{Q.16} & \multicolumn{3}{l}{$\lluntil{\neg \mathtt{ sampleA }} {(\lcall \land \mathtt{sampleA} \land \lcdnext (\lqry \land \ldnext (\lcall \land \mathtt{sampleA} \land \lcdnext (\lqry \land \ldnext (\lcall \land \mathtt{sampleA})))))}$} \\
    \textbf{S.1}  & \multicolumn{3}{l}{$\llglob (\lcall \land \mathtt{B} \land \mathtt{sorted} \land \mathtt{valOccurs} \land  \mathtt{left} \leq \mathtt{right} \implies \lcunext \mathtt{a[mid]} == \mathtt{val})$} \\
    \textbf{S.2} & \multicolumn{3}{l}{$\llglob (\lcall \land \mathtt{B} \land \mathtt{sorted} \land \neg \mathtt{valOccurs} \land  \mathtt{left} \leq \mathtt{right} \implies \lcunext \mathtt{a[mid]} \neq \mathtt{val})$} \\
    \textbf{S.3} & \multicolumn{3}{l}{$\llglob (\lcall \land \mathtt{B} \land \mathtt{sorted} \land \mathtt{valOccurs} \land  \mathtt{left} < \mathtt{right} \implies \ldnext \lleven (\lcall \land \mathtt{B}))$} \\
    \textbf{S.4} & \multicolumn{3}{l}{$\llglob (\lcall \land \mathtt{B} \land \mathtt{sorted} \land \neg \mathtt{valOccurs} \land  \mathtt{left} < \mathtt{right} \implies \ldnext \lleven (\lcall \land \mathtt{B}))$} \\
    \bottomrule
    \end{tabular}
    \end{adjustbox}
\end{table}

\slimparagraph{Setup.}
Our benchmark consists of three programs (\emph{\textbf{Schelling}}, \emph{\textbf{Tic-tac-toe}} and \emph{\textbf{Virus}}) and 16 mixed LTL/\acs{POTLF} formulae (Table \ref{tab:formulae_experiments}, \textbf{Q} formulae).
All programs consist of a potentially unbounded sequence of nested queries.
This places them at the frontier of probabilistic programming: even solving simple questions about their posterior distribution is a hardly tractable problem~\cite{Rainforth18}. Additionally, to inquire scalability, we provide a case study on the \emph{\textbf{Sherwood}} binary search with increasing program state space, against four formulae (Table \ref{tab:formulae_experiments}, \textbf{S} formulae).
We run the experiments on a machine with a 4.5GHz 8-core AMD CPU and 64~GB of RAM running Ubuntu 24.04.
We do not offer an experimental comparison with PReMo~\cite{WojtczakE07} nor Pray~\cite{WinklerK23a}, since they only compute termination probabilities on \acp{pPDA}.
This task is not the bottleneck of the overall model checking algorithm---our tool always computes them in less than one second.

We describe in the following the three programs, (some of) the formulae we verify on them, and their results. Afterwards, we present the case study separately.
Two formulae inspecting conditioning are of interest for all three
(\textbf{Q.1} and \textbf{Q.2}). 
If an observe statement conditions on an event with zero probability, 
all runs of a query are rejected. 
Such \emph{ill-defined} queries~\cite{Jacobs21} 
do not represent a valid probability distribution, which is undesired.
\acs{LTL} property \textbf{Q.1} means that a run is always sooner or later reinstantiated into a feasible one.
However, if queries indefinitely call each other in a nonterminating program, a diverging run 
indefinitely hits unsatisfied observations in the infinite nesting.
\acs{POTLF} formula \textbf{Q.2} means that no procedure is indefinitely reinstantiated by an observation: either it terminates, or it diverges.

\begin{wrapfigure}[16]{r}{0.35\textwidth}
\centering
\vspace{-3em}
\begin{minted}[fontsize=\scriptsize]{python}
u4 p;
main() {
  bool res;
  p = 0{2:6}1{1:6}2{1:6}3{1:6}4;
  query alice(res);
}
Alice(bool &res) {
  aliceLoc = Bernoulli(55,100);
  query Bob(bobLoc);
  observe (aliceLoc == bobLoc);
  res = aliceLoc;
}
Bob(bool &res) {
  bobLoc = Bernoulli(55,100);
  fair = true {p:10} false;
  if (fair) {
    query Alice(aliceLoc);
    observe (bobLoc == aliceLoc);
  } 
  res = bobLoc;
}
\end{minted}
\vspace{-1em}
\caption{Schelling.}
\vspace{-1em}
\label{fig:coordination_game}
\end{wrapfigure}
\slimparagraph{\textbf{Schelling}.}
We consider an instance 
of a \emph{Schelling coordination game}~\cite{StuhlmullerG14,Schelling1980}.
Two agents wish to meet in town but cannot communicate.
However, they know perfectly each other's preferences.
Each agent samples both a location according to its preference, and one obtained simulating the behavior of the other agent. Finally, it conditions on the two being equal.
The original program~\cite{StuhlmullerG14} bounds the recursion 
depth by a constant.
Conversely, we allow for unbounded recursion,  but we introduce 
a global variable \texttt{p} controlling the recursion probability. 
We fix \texttt{p}'s distribution so that the program is \textsc{ast}.
Formula \textbf{Q.13} is the event that the first call to \texttt{Bob} is rejected,
while \textbf{Q.14} that the first call to \texttt{Alice} is \emph{not} rejected. Since 
Bob is not always fair, his procedure is much less likely to be rejected.

\slimparagraph{\textbf{Tic-tac-toe}}.
Recursive probabilistic programs also
model multi-player games involving sequential decision-making, i.e., where the best move in a turn depends on moves in subsequent turns.
In these games, players choose actions 
\begin{wrapfigure}[5]{r}{0.15\textwidth}
\vspace{-5ex}
\centering
\large
\begin{tabular}{c|c|c}   & $\ocircle$ &  \\      \hline
 $\ocircle$  & $\times$ & $\times$ \\      \hline
   & $\ocircle$ &
\end{tabular}
\vspace{-2ex}
\caption{~\cite{StuhlmullerG14}}
\label{fig:tic-tac-toe}
\end{wrapfigure}
that maximize future rewards by simulating, through nested queries,
other players' turns as the game progresses.
We consider tic-tac-toe~\cite{StuhlmullerG14}. 
A procedure modelling a
player's reasoning first marks uniformly at random a cell between those not taken yet; then it recursively queries itself to simulate the other player's behaviour with the updated grid. This way, the whole game is recursively simulated, ending in a draw or in one of the two players winning. Finally, it conditions on a coin flip with weight corresponding to the utility of the outcome, thus performing softmax optimal decision making.
We inspect the max/min number of turns until termination of the game,
corresponding to the program's max/min recursion depth.
\textbf{Q.4} means that there are at least two recursive calls (i.e., two turns)%
---what we would expect from the initial state of Fig.~\ref{fig:tic-tac-toe},
where no move can immediately terminate the game.
We also examine the players' meta-reasoning:
\textbf{Q.5} excludes that a player ever thinks
that the other one will pick cell $[0,1]$ in the next turn, which is indeed already marked at the beginning.

\begin{wrapfigure}[16]{r}{0.33\textwidth}
\centering
\vspace{-3em}
\begin{minted}[fontsize=\scriptsize]{python}
young() {
   bool f;
   … (other statements) 
   … (sample infecting young)
   if (infect_young) {
     query young();
   }
   … (sample infecting elders)
   if (infect_elder) {
     query elder(&f);
     query elder(&f);
   }
}
elder(bool &f) {
   … (same as young())
   … (sample passing away)
   if (pass_away) {
        f = 1;
   } else { f = 0; }
}
\end{minted}
\vspace{-1em}
\caption{Virus (sketch).}
\vspace{-1em}
\label{fig:virus}
\end{wrapfigure}
\slimparagraph{\textbf{Virus}}.
Recursive probabilistic programs encode epidemiological models
and general multi-type Branching Processes~\cite{EtessamiY05b,haccou2005branching} as models of population dynamics in biology.
They consist of an unbounded population of susceptible individuals
belonging to distinct species.
In the programs, the stack models the individuals currently involved in the system, and 
different procedures model different species' behaviors.
We consider the program from \cite[Fig.~1]{WinklerGK22}: a virus outbreak. Our population is composed of either young or elder individuals. 
We introduce conditioning in the model expressing that young ones are less likely to transmit the virus.
\textbf{Q.6} encodes the CaRet formula of~\cite[p.~3]{WinklerGK22}:
a chain of infections of young people leads to the death of an elder person.
In \acs{POTLF}, $\lcduntil{(\neg \mathtt{elder})}{(\mathtt{young} \land f)}$, where $f$ is a boolean variable representing an elder person passing away. Young people cannot pass away in our model.
Note here the need for the $\chi$ variant of the $\lluntil{}{}$ operator to capture exactly the property.
Consider the case of a young person infecting first an elder person who survives, and then an elder who passes away.
There would be only one young person in the chain of infections, however the LTL $\lluntil{(\neg \mathtt{elder})}{(\mathtt{young} \land f)}$ would not hold: the first elder's infection happens temporally before the death, although not part of the chain.
\acs{POTLF} instead captures this context-free property: $\lcduntil{}{}$ exploits the $\chi$ relation to skip the first call to $\texttt{elder()}$---this example trace would satisfy it.
\textbf{Q.7} means that eventually an elder person infects at least one elder person,
but none of them passes away.

\slimparagraph{Results.}
We report the experimental results for building the support chain (Table~\ref{tab:termResults}, cf. Fig.~\ref{fig:workflow-diagram}) and for solving qualitative and quantitative model checking (Table~\ref{tab:results}, cf. Fig.~\ref{fig:qualitative-workflow-diagram}). All times are in seconds and the timeout is 1 h.
A key metric is the number of equations arising from each instance.
Since we decompose the graph into SCCs and solve it bottom-up, many equations are solved just by propagating already computed values.
We thus report how many equations actually require running our semi-algorithm because they introduce circular dependency between variables in the system.
We call them \textbf{non-trivial} (NT) equations. 
In Table~\ref{tab:termResults}, 
$|Q_\mathcal{A}|$,$|f|_{(\mathcal{A})}$, $|f_{NT}|{_{(\mathcal{A})}}$ 
are the number of resp.\ states, equations and non-trivial equations
that arise when analysing \ac{pOPA} $\mathcal{A}$ modeling the program. 
To estimate the size of individual SCCs we feed to OVI/Z3,
we report also $|f_{NT}(\text{SCC})|{_\mathit{max}}_{(\mathcal{A})}$, 
the maximum number of non-trivial equations 
in an SCC of the equation system of $\mathcal{A}$. 
In Table~\ref{tab:results}, AS tells whether the formula holds almost surely. 
If it is not the case, we perform quantitative model checking to get probability P, 
for which we provide the same metrics as above, 
but on $\hat{\mathcal{A}}$, the cross-product between $\mathcal{A}$ 
and the formula automaton (Sec.~\ref{sec:mc-algorithm-separation}).

\slimparagraph{\textbf{Sherwood}.}
The \acs{pOPA} state space is ruled by two parameters: 
\texttt{K}, the number of bits of each array element, 
and \texttt{M}, the array's size. We investigate \texttt{K} $\in [1,4]$ and \texttt{M} $\in [1,7]$.
For each instance, the main procedure randomly initializes the array and a value to search, and calls \B. We assume that calls to \texttt{uniform()} diverge if \texttt{left} $>$ \texttt{right}.
Formula \textbf{S.1} is partial correctness from Sec.~\ref{sec:intro}. 
Formula \textbf{S.2} is a dual version. 
Formulae \textbf{S.3} and \textbf{S.4} are stack-inspection properties: 
if \B{} is invoked in a state where \texttt{left} $\leq$ \texttt{right}, 
\arr{} is sorted and \texttt{val} occurs in \arr{} (does not, in S.4), 
then there is a recursive call to \B{}. All formulae hold almost surely except S.3: 
if the array has multiple elements, the randomized pivot selection may pick the searched value, terminating the program immediately.
Fig.~10 and Table 2 show results for qualitative model checking on all formulae, and quantitative for some parameters of S.3.

\begin{figure}[bt]
\begin{minipage}{.6\textwidth}
  \centering
  \includegraphics[width=\textwidth]{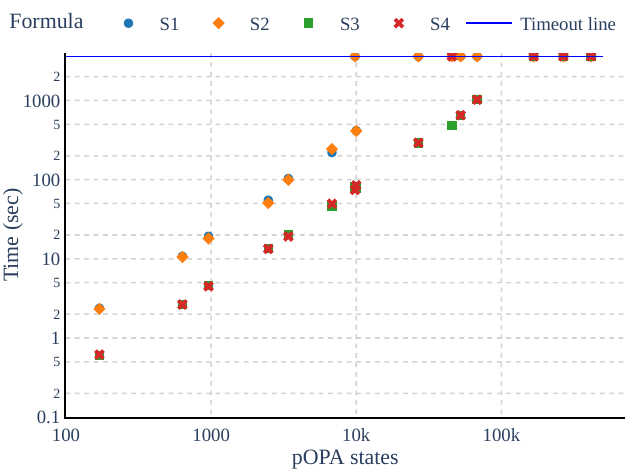}
\end{minipage}%
\begin{minipage}{.4\textwidth}
    \centering
    \scriptsize
    \begin{tabular}{c c @{\hspace{.5em}} | @{\hspace{.5em}} r @{\hspace{.5em}} r @{\hspace{.5em}} r }
    \toprule
    &&\multicolumn{3}{c}{Quantitative MC} \\
    \midrule
    K & M & $|f|{_{(\hat{\mathcal{A}})}}$ & $t_\mathit{tot}$ & P\\
    \midrule
          1  &  1  & 154k   & 1.18    & 1  \\
          1  &  2  & 2 M    & 12.12  & $\sim$ 0.906  \\
          1  &  3  & 8 M    & 53.87  & $\sim$ 0.882  \\
          2  &  1  & 598k   & 5.20   & 1  \\
           2 &  2  & 15 M   & 111.36  & $\sim$ 0.961 \\
           3 &  1  & 2 M  & 23.24  & 1  \\
           4 &  1  & 9 M  & 123.63 & 1  \\
    \bottomrule
    \end{tabular}
\end{minipage}
    \vspace{-2ex}
    \captionlistentry[table]{Qualitative model checking of all formulae (left) and quantitative model checking}
    \captionsetup{labelformat=andtable}
    \caption{Sherwood binary search: qualitative model checking of all formulae (left, both axis are on logarithmic scale) and quantitative model checking of S.3 for some values of \texttt{K} and \texttt{M} (right).}
    \label{tab:results-sherwood}
\end{figure}

\begin{table}[bt]
\caption{Experimental results for the computation of termination probabilities.
  $t_{Z3}$ (resp. $t_{OVI}$) is the time required to compute upper bounds with Z3 (OVI).
}
\label{tab:termResults}
\centering
\begin{adjustbox}{max width=\textwidth}
\renewcommand{\arraystretch}{1.1}
\begin{tabular}{l @{\hspace{.5em}} | @{\hspace{.5em}} r @{\hspace{.5em}} r @{\hspace{.5em}} r @{\hspace{.5em}} r @{\hspace{.5em}} | @{\hspace{.5em}} r @{\hspace{.5em}} r @{\hspace{.5em}} | r}
\toprule
name & $|Q_\mathcal{A}|$ &  $|f|_{(\mathcal{A})}$  & $|f_{NT}|{_{(\mathcal{A})}}$ & $|f_{NT}(\text{SCC})|{_\mathit{max}}_{(\mathcal{A})}$ & $t_\mathit{Z3}$ & $t_\mathit{OVI}$ & $t_\mathit{PAST}$ \\
\midrule
Schelling   & 311   & 1230  & 266 & 52 & TO   &  0.03  &  0.03  \\
Tic-tac-toe &  1780 & 3443 & 248 & 36 & 0.02 &  0.01  &  0.08 \\
Virus       &  427  & 21211 & 6490 & 6488 & TO   &   0.38 &  0 \\
\bottomrule
\end{tabular}
\end{adjustbox}
\end{table}

\begin{table}[bt]
\caption{Qualitative and quantitative MC.
  $t_\mathit{G}$, $t_\mathit{OVI}$ are times spent resp. analyzing the synchronized product and computing with OVI upper bounds for non trivial equations.
}
\label{tab:results}
\centering
\scriptsize
\begin{adjustbox}{max width=\textwidth}
\scriptsize
\begin{tabular}{l c | r r r @{\hspace{.5em}} | r r r r r c }
\toprule
\multicolumn{2}{c}{} & 
\multicolumn{3}{|c|}{Qualitative MC} &
\multicolumn{6}{c}{Quantitative MC} \\
\midrule
name & $\varphi$ & $t_G$ & $t_\mathit{tot}$ & AS & $|f|{_{(\hat{\mathcal{A}})}}$ & $|f_{NT}|{_{(\hat{\mathcal{A}})}}$ & $|f_{NT}(\text{SCC})|{_\mathit{max}}_{(\hat{\mathcal{A}})}$ & $t_\mathit{OVI}$ & $t_\mathit{tot}$ & P\\
\midrule
\parbox[t]{3em}{\centering\multirow{8}{*}{\rotatebox[origin=c]{90}{Schelling}}} &
    Q.1   & 0.02   & 0.15   &  \cmark  & -  & -  & -  & -  & -  &  1\\
&    Q.2  & 26.28  & 26.40  &  \cmark  & -  & -  & -  & -  & -  &  1 \\
&    Q.3  & 1.25   & 1.37   &  \xmark  & 893k  & 68k  &  7.3k  & 11.71 & 20.58  &   $\sim$ 0.895 \\
&    Q.8  & 0.22  & 0.34   &  \xmark  & 214k   & 15k  & 1.4k   & 3.80 & 5.70  &  $\sim$ 0.610 \\
&    Q.9 & 0.04  & 0.16   &  \xmark  & 36k   & 3.2k  & 252 & 1.29  & 1.57  &  $\sim$ 0.610 \\
&    Q.12 & 1.01  & 1.14   &  \xmark  & 651k   & 61k  & 6.7k  & 11.56 & 20.10  & $\sim$ 0.096  \\
&    Q.13 & 0.67  & 0.80   &  \xmark  & 599k   & 43k & 5.7k   & 6.48 & 13.28 & $\sim$ 0.506 \\
&    Q.15 & 0.95 & 1.07   &  \xmark  & 762k  & 63k  & 7.5k & 12.49 &  23.21 & $\sim$ 0.543 \\

\addlinespace[2pt]
\midrule
\parbox[t]{3em}{\centering\multirow{8}{*}{\rotatebox[origin=c]{90}{Tic-tac-toe}}} &
    Q.1    & 0.12   & 0.2    & \cmark  & -  & -  & - & - & -  &  1\\
&   Q.2    & 122.16  & 122.33  & \cmark  & -  & -  & - & - & -  &  1\\
&   Q.4    & 9.79  & 9.98  & \cmark  & -  & -  & -  & - & - &  1\\
&   Q.5    & 8.22   & 8.41   & \cmark  & -  & -  & -  & - & -  &  1\\
&   Q.10   & 2.94    & 3.10    & \xmark  & 1.1 M & 77k & 3k & 0.79 & 10.45 & $\sim$ 0.712 \\
&   Q.11    & 2.96    & 3.13    & \cmark  & - & -  & -  & - & - &  1 \\
&   Q.14   & 0.54    & 0.71    & \xmark  & 280k & 20k & 1.5k & 0.17 & 2.30 & $\sim$ 0.288  \\
&   Q.16    & 386.10    & 386.22    & \cmark  & -  & -  & - & -  & -  &  1 \\
\addlinespace[2pt]
\midrule
\parbox[t]{3em}{\centering\multirow{4}{*}{\rotatebox[origin=c]{90}{Virus}}} &
    Q.1 & 0.18   & 2.13   & \xmark  & 243k & 25k & 12k & 0.50 & 9.72 & $\sim$ 0.239 \\
&   Q.2 & 263.94 & 265.9  & \cmark  & -  & -   & -  & - & - &  1\\
&   Q.6 & 14.96  & 16.95   & \xmark  & ?  & ?   & ? & ? &  ? & TO  \\
&   Q.7 & 891.45  & 893.37    & \xmark  & ?   & ?  & ? & ? & ? &  TO \\
\bottomrule
\end{tabular}
\end{adjustbox}
\end{table}

\section{Discussion and Future Work}
Experimental results show that OVI scales better than Z3 in solving systems for building the support chain. Z3 already times out when provided with a system of 52 equations.
For quantitative model checking, the exponential blow-up on the formula automaton generates in our benchmark hundred of thousands of equations. This suggests that determinization-based approaches, leading to another exponentiation, would not work. 
Finally, our tool scales up to a few million of equations, supporting checking non-trivial formulae on medium-size \acp{pPDA}.

Future improvements may come from finding ways of modeling non-trivial probabilistic programs with less expressive \ac{pPDA} subclasses that have more efficient model checking algorithms.
\acp{pPDA} with one single state (or \emph{stochastic context-free grammars})
admit P-time algorithms for computing the probability of termination \cite{EtessamiSY12};
\emph{probabilistic one-counter automata}
admit P-time algorithms for model checking $\omega$-regular properties \cite{StewartEY15}.


\begin{credits}
\subsubsection{\ackname}
\sloppy
We thank Tobias Winkler and Joost-Pieter Katoen (RWTH Aachen)
for the fruitful discussions and advice on implementing OVI.
This work was funded by the Vienna Science and Technology Fund (WWTF)
grants [10.47379/ICT19018] (ProbInG) and ICT22-023 (TAIGER),
and by the Horizon Europe programme
grants No.\ 101034440 (MSCA Doctoral Network LogiCS@TU Wien)
and No.\ 101107303 (MSCA Postdoctoral Fellowship CORPORA) \includegraphics[height=\fontcharht\font`\B]{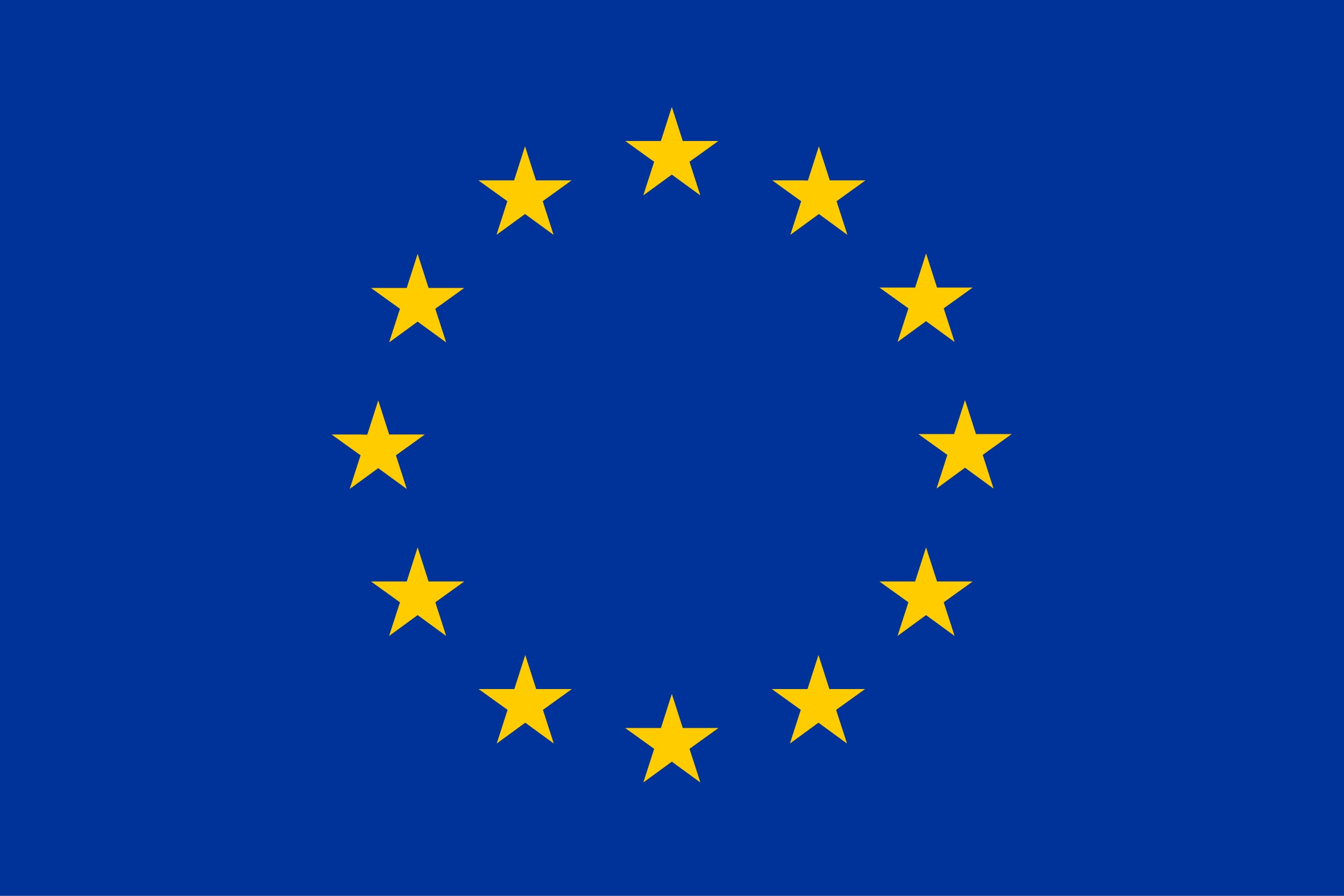}.

\subsubsection{\discintname}
The authors declare no competing interests for this article.
\end{credits}

%
%
%
%
%
%
\bibliographystyle{splncs04}
\bibliography{biblio,prob}

\begin{thebibliography}{10}
\providecommand{\url}[1]{\texttt{#1}}
\providecommand{\urlprefix}{URL }
\providecommand{\doi}[1]{https://doi.org/#1}

\bibitem{BaierK08}
Baier, C., Katoen, J.: Principles of Model Checking. {MIT} Press (2008)

\bibitem{BaierK00023}
Baier, C., Kiefer, S., Klein, J., M{\"{u}}ller, D., Worrell, J.: Markov chains and unambiguous automata. J. Comput. Syst. Sci.  \textbf{136},  113--134 (2023). \doi{10.1016/J.JCSS.2023.03.005}

\bibitem{BrazdilBHK08}
Br{\'{a}}zdil, T., Brozek, V., Holecek, J., Kucera, A.: Discounted properties of probabilistic pushdown automata. In: {LPAR}'08. LNCS, vol.~5330, pp. 230--242. Springer (2008). \doi{10.1007/978-3-540-89439-1\_17}

\bibitem{BrazdilEKK13}
Br{\'{a}}zdil, T., Esparza, J., Kiefer, S., Kucera, A.: Analyzing probabilistic pushdown automata. Formal Methods Syst. Des.  \textbf{43}(2),  124--163 (2013). \doi{10.1007/s10703-012-0166-0}

\bibitem{BrazdilEK05}
Br{\'{a}}zdil, T., Esparza, J., Kucera, A.: Analysis and prediction of the long-run behavior of probabilistic sequential programs with recursion (extended abstract). In: {FOCS} '05. pp. 521--530. {IEEE} Computer Society (2005). \doi{10.1109/SFCS.2005.19}

\bibitem{BrazdilKKV15}
Br{\'{a}}zdil, T., Kiefer, S., Kucera, A., Varekov{\'{a}}, I.H.: Runtime analysis of probabilistic programs with unbounded recursion. J. Comput. Syst. Sci.  \textbf{81}(1),  288--310 (2015). \doi{10.1016/J.JCSS.2014.06.005}

\bibitem{Canny88}
Canny, J.F.: Some algebraic and geometric computations in {PSPACE}. In: {STOC}'88. pp. 460--467. {ACM} (1988). \doi{10.1145/62212.62257}

\bibitem{ChiariMPP23}
Chiari, M., Mandrioli, D., Pontiggia, F., Pradella, M.: A model checker for operator precedence languages. ACM Trans. Program. Lang. Syst.  \textbf{45}(3) (2023). \doi{10.1145/3608443}

\bibitem{ChiariMP21a}
Chiari, M., Mandrioli, D., Pradella, M.: Model-checking structured context-free languages. In: {CAV}'21. LNCS, vol. 12760, p. 387–410. Springer (2021). \doi{10.1007/978-3-030-81688-9_18}

\bibitem{ChiariMP21b}
Chiari, M., Mandrioli, D., Pradella, M.: A first-order complete temporal logic for structured context-free languages. Log. Methods Comput. Sci.  \textbf{18:3} (2022). \doi{10.46298/LMCS-18(3:11)2022}

\bibitem{CouvreurSS03}
Couvreur, J., Saheb, N., Sutre, G.: An optimal automata approach to {LTL} model checking of probabilistic systems. In: {LPAR}'03. LNCS, vol.~2850, pp. 361--375. Springer (2003). \doi{10.1007/978-3-540-39813-4\_26}

\bibitem{EsparzaKL10}
Esparza, J., Kiefer, S., Luttenberger, M.: Computing the least fixed point of positive polynomial systems. {SIAM} J. Comput.  \textbf{39}(6),  2282--2335 (2010). \doi{10.1137/090749591}

\bibitem{EsparzaKS20}
Esparza, J., Kret{\'{\i}}nsk{\'{y}}, J., Sickert, S.: A unified translation of linear temporal logic to {\(\omega\)}-automata. J. {ACM}  \textbf{67}(6),  33:1--33:61 (2020). \doi{10.1145/3417995}

\bibitem{EsparzaKM04}
Esparza, J., Kucera, A., Mayr, R.: Model checking probabilistic pushdown automata. In: {LICS}'04. pp. 12--21. {IEEE} Computer Society (2004). \doi{10.1109/LICS.2004.1319596}

\bibitem{EtessamiSY12}
Etessami, K., Stewart, A., Yannakakis, M.: Polynomial time algorithms for multi-type branching processesand stochastic context-free grammars. In: {STOC}'12. pp. 579--588. {ACM} (2012). \doi{10.1145/2213977.2214030}

\bibitem{EtessamiY05}
Etessami, K., Yannakakis, M.: Algorithmic verification of recursive probabilistic state machines. In: {TACAS}'05. LNCS, vol.~3440, pp. 253--270. Springer (2005). \doi{10.1007/978-3-540-31980-1\_17}

\bibitem{EtessamiY05b}
Etessami, K., Yannakakis, M.: Recursive markov chains, stochastic grammars, and monotone systems of nonlinear equations. In: {STACS}'05. LNCS, vol.~3404, pp. 340--352. Springer (2005). \doi{10.1007/978-3-540-31856-9\_28}

\bibitem{EtessamiY09}
Etessami, K., Yannakakis, M.: Recursive markov chains, stochastic grammars, and monotone systems of nonlinear equations. J. {ACM}  \textbf{56}(1),  1:1--1:66 (2009). \doi{10.1145/1462153.1462154}

\bibitem{EtessamiY12}
Etessami, K., Yannakakis, M.: Model checking of recursive probabilistic systems. {ACM} Trans. Comput. Log.  \textbf{13}(2),  12:1--12:40 (2012). \doi{10.1145/2159531.2159534}

\bibitem{EtessamiY15}
Etessami, K., Yannakakis, M.: Recursive markov decision processes and recursive stochastic games. J. {ACM}  \textbf{62}(2),  11:1--11:69 (2015). \doi{10.1145/2699431}

\bibitem{Evans17}
Evans, O., Stuhlm\"{u}ller, A., Salvatier, J., Filan, D.: Modeling agents with probabilistic programs (2017), \url{http://agentmodels.org}

\bibitem{Floyd1963}
Floyd, R.W.: Syntactic analysis and operator precedence. J. {ACM}  \textbf{10}(3),  316--333 (1963). \doi{10.1145/321172.321179}

\bibitem{Ghahramani15}
Ghahramani, Z.: Probabilistic machine learning and artificial intelligence. Nat.  \textbf{521}(7553),  452--459 (2015). \doi{10.1038/NATURE14541}

\bibitem{GoodmanMRBT08}
Goodman, N.D., Mansinghka, V.K., Roy, D.M., Bonawitz, K.A., Tenenbaum, J.B.: Church: a language for generative models. In: {UAI}'08. pp. 220--229. {AUAI} Press (2008)

\bibitem{probmods2}
Goodman, N.D., Tenenbaum, J.B., {The ProbMods Contributors}: {Probabilistic Models of Cognition}. \url{http://probmods.org/v2} (2016), accessed: 2025-5-22

\bibitem{GordonHNR14}
Gordon, A.D., Henzinger, T.A., Nori, A.V., Rajamani, S.K.: Probabilistic programming. In: {FOSE}'14. pp. 167--181. {ACM} (2014). \doi{10.1145/2593882.2593900}

\bibitem{haccou2005branching}
Haccou, P., Jagers, P., Vatutin, V.A.: Branching processes: variation, growth, and extinction of populations. Cambridge University Press (2005)

\bibitem{HenselJKQV22}
Hensel, C., Junges, S., Katoen, J., Quatmann, T., Volk, M.: The probabilistic model checker {Storm}. Int. J. Softw. Tools Technol. Transf.  \textbf{24}(4),  589--610 (2022). \doi{10.1007/s10009-021-00633-z}

\bibitem{Jacobs21}
Jacobs, J.: Paradoxes of probabilistic programming: and how to condition on events of measure zero with infinitesimal probabilities. {ACM} Program. Lang.  \textbf{5}({POPL}),  1--26 (2021). \doi{10.1145/3434339}

\bibitem{JovanovicM12}
Jovanovic, D., de~Moura, L.: Solving non-linear arithmetic. {ACM} Commun. Comput. Algebra  \textbf{46}(3/4),  104--105 (2012). \doi{10.1145/2429135.2429155}

\bibitem{KuceraEM06}
Kucera, A., Esparza, J., Mayr, R.: Model checking probabilistic pushdown automata. Log. Methods Comput. Sci.  \textbf{2}(1) (2006). \doi{10.2168/LMCS-2(1:2)2006}

\bibitem{KwiatkowskaNP11}
Kwiatkowska, M.Z., Norman, G., Parker, D.: {PRISM} 4.0: Verification of probabilistic real-time systems. In: {CAV}'11. LNCS, vol.~6806, pp. 585--591. Springer (2011). \doi{10.1007/978-3-642-22110-1\_47}

\bibitem{MP18}
Mandrioli, D., Pradella, M.: Generalizing input-driven languages: Theoretical and practical benefits. Computer Science Review  \textbf{27},  61--87 (2018). \doi{10.1016/j.cosrev.2017.12.001}

\bibitem{mcconnell2007analysis}
McConnell, J.: Analysis of algorithms. Jones \& Bartlett Publishers (2007)

\bibitem{MouraB08}
de~Moura, L.M., Bj{\o}rner, N.S.: {Z3:} an efficient {SMT} solver. In: {TACAS} 2008. LNCS, vol.~4963, pp. 337--340. Springer (2008). \doi{10.1007/978-3-540-78800-3\_24}

\bibitem{OlmedoGJKKM18}
Olmedo, F., Gretz, F., Jansen, N., Kaminski, B.L., Katoen, J., McIver, A.: Conditioning in probabilistic programming. {ACM} Trans. Program. Lang. Syst.  \textbf{40}(1),  4:1--4:50 (2018). \doi{10.1145/3156018}

\bibitem{OlmedoKKM16}
Olmedo, F., Kaminski, B.L., Katoen, J., Matheja, C.: Reasoning about recursive probabilistic programs. In: {LICS}'16. pp. 672--681. {ACM} (2016). \doi{10.1145/2933575.2935317}

\bibitem{Pnueli77}
Pnueli, A.: The temporal logic of programs. In: FOCS '77. pp. 46--57. {IEEE} Computer Society (1977). \doi{10.1109/SFCS.1977.32}

\bibitem{Pontiggia21}
Pontiggia, F.: {POMC}. {A} model checking tool for operator precedence languages on omega-words. Master's thesis, Politecnico di Milano (2021), \url{http://hdl.handle.net/10589/176028}

\bibitem{abs-2404-03515}
Pontiggia, F., Bartocci, E., Chiari, M.: Model checking probabilistic operator precedence automata. CoRR  (2024), \url{https://doi.org/10.48550/arXiv.2404.03515}

\bibitem{PontiggiaCP21}
Pontiggia, F., Chiari, M., Pradella, M.: Verification of programs with exceptions through operator precedence automata. In: SEFM'21. LNCS, vol. 13085, pp. 293--311. Springer, Berlin, Heidelberg (2021). \doi{10.1007/978-3-030-92124-8\_17}

\bibitem{Rainforth18}
Rainforth, T.: Nesting probabilistic programs. In: Globerson, A., Silva, R. (eds.) {UAI} '18. pp. 249--258. {AUAI} Press (2018), \url{http://auai.org/uai2018/proceedings/papers/92.pdf}

\bibitem{Renegar92}
Renegar, J.: On the computational complexity and geometry of the first-order theory of the reals, parts {I}--{III}. J. Symb. Comput.  \textbf{13}(3),  255--352 (1992). \doi{10.1016/S0747-7171(10)80003-3}

\bibitem{Schelling1980}
Schelling, T.C.: {The Strategy of Conflict}. Harvard University Press (1980)

\bibitem{speechact}
Scontras, G., Tessler, M.H., Franke, M.: {Probabilistic language understanding: An introduction to the Rational Speech Act framework}. \url{https://www.problang.org/}, accessed: 2025-5-22

\bibitem{seaman2020}
Seaman, I.R., van~de Meent, J.W., Wingate, D.: Nested reasoning about autonomous agents using probabilistic programs (2020), \url{https://arxiv.org/abs/1812.01569}

\bibitem{StewartEY15}
Stewart, A., Etessami, K., Yannakakis, M.: Upper bounds for {Newton}'s method on monotone polynomial systems, and {P}-time model checking of probabilistic one-counter automata. J. {ACM}  \textbf{62}(4),  30:1--30:33 (2015). \doi{10.1145/2789208}

\bibitem{StuhlmullerG14}
Stuhlm{\"{u}}ller, A., Goodman, N.D.: Reasoning about reasoning by nested conditioning: Modeling theory of mind with probabilistic programs. Cognitive Systems Research  \textbf{28},  80--99 (2014). \doi{10.1016/J.COGSYS.2013.07.003}

\bibitem{VardiW94}
Vardi, M.Y., Wolper, P.: Reasoning about infinite computations. Inf. Comput.  \textbf{115}(1),  1--37 (1994). \doi{10.1006/INCO.1994.1092}

\bibitem{WinklerGK22}
Winkler, T., Gehnen, C., Katoen, J.: Model checking temporal properties of recursive probabilistic programs. In: {FOSSACS}'22. LNCS, vol. 13242, pp. 449--469. Springer (2022). \doi{10.1007/978-3-030-99253-8\_23}

\bibitem{WinklerK23a}
Winkler, T., Katoen, J.: Certificates for probabilistic pushdown automata via optimistic value iteration. In: {TACAS}'23. LNCS, vol. 13994, pp. 391--409. Springer (2023). \doi{10.1007/978-3-031-30820-8\_24}

\bibitem{WinklerK23b}
Winkler, T., Katoen, J.: On certificates, expected runtimes, and termination in probabilistic pushdown automata. In: {LICS}'23. pp. 1--13 (2023). \doi{10.1109/LICS56636.2023.10175714}

\bibitem{WojtczakE07}
Wojtczak, D., Etessami, K.: {PReMo}: An analyzer for probabilistic recursive models. In: {TACAS}'07. LNCS, vol.~4424, pp. 66--71. Springer (2007). \doi{10.1007/978-3-540-71209-1\_7}

\bibitem{YannakakisE05}
Yannakakis, M., Etessami, K.: Checking {LTL} properties of recursive {Markov} chains. In: {QEST}'05. pp. 155--165. {IEEE} (2005). \doi{10.1109/QEST.2005.8}

\bibitem{ZhangA22}
Zhang, Y., Amin, N.: Reasoning about ``reasoning about reasoning'': semantics and contextual equivalence for probabilistic programs with nested queries and recursion. {ACM} Program. Lang.  \textbf{6}({POPL}),  1--28 (2022). \doi{10.1145/3498677}

\end{thebibliography}


\end{document}